\documentstyle[12pt,aas2pp4]{article}
\def\insfig#1{#1}
\def\endinsfig{\end{document}}

\font\smallrm=cmr8

\def\FWHM{{\smallrm FWHM}}

\def\kms{\hbox{$\,$km$\,$s$^{-1}$}}
\def\kmsMpc{\hbox{$\,$km$\,$s$^{-1}\,$Mpc$^{-1}$}}

\lefthead{Tonry}
\righthead{Gravitational Lens Redshifts}

\begin{document}

\title{Redshifts of the Gravitational Lenses B1422+231 and PG1115+080\altaffilmark{1}}

\author{John L. Tonry}
\affil{Institute for Astronomy, University of Hawaii, Honolulu, HI 96822}
\affil{Electronic mail: jt@avidya.ifa.hawaii.edu}
\authoremail{jt@avidya.ifa.hawaii.edu}

\altaffiltext{1}{Based on observations at the W. M. Keck Observatory,
which is operated jointly by the California Institute of Technology
and the University of California}

\begin{abstract}
B1422+231 and PG1115+080 are gravitational lens systems producing
quadruple QSO images where there is real promise that time delays can
constrain the Hubble constant.  In addition, the lensing galaxies are
both part of groups which can play an important role in
modelling the lens potential.  This article reports redshifts for the
lensing galaxies and three neighboring galaxies in each of the two
systems.  B1422+231 consists of a group at $z=0.339$ with a dispersion
of 733\kms, and PG1115+080 is a group at $z=0.311$ with a dispersion
of 326\kms.  One of the neighboring galaxies in the B1422+231 system
turned out to be an emission line galaxy at $z=0.536$, suggesting that
QSO light passing through B1422+231 may have been subjected to lensing
by a cluster at this more distant redshift.  The velocity dispersion
of the lensing galaxy in PG1115+080 is determined to be
$281\pm25$\kms\ (1\arcsec\ square aperture), which is surprisingly
large given the image splittings of 1.2\arcsec\ in that system.

\noindent \it{Subject headings:} cosmology --- distance scale ---
gravitational lensing --- quasars: individual (B1422+231, PG1115+080)

\end{abstract}

\section{Introduction}

The differing paths followed by light around a gravitational lens
leads to different time of flight.  If the lensed source should vary,
measurement of a time delay can transform dimensionless redshifts into
physical distances, hence provide a Hubble constant (Refsdal 1964).
To within factors of order unity the entire time of flight is simply
the time delay divided by the square of the image splitting (in
radians).  In principle this is an extremely powerful method to learn
about cosmology; in practice it has proven to be difficult to measure
time delays and model lenses accurately enough to challenge the 
accuracy of 10--20 percent claimed by more traditional measures of
$H_0$.  Nevertheless, gravitational lens measurements of cosmology are
critically important because they are potentially subject to far fewer
biases than the usual distance ladder.

Time delays have now been measured for two lens systems: 0957+561 and
PG1115+080.  The difficulty in measuring a time delay and then
rendering from it cosmological information is illustrated by 0957+561.
A time delay was reported by Schild and Cholfin (1986), disputed by Press,
Rybicki, and Hewitt (1992), and finally resolved in favor of Schild's
number by further observations by Kundi\'c et al. (1996).  Models of
the system have also undergone improvement, with estimated values for
$H_0$ as low as 50\kmsMpc\ and as high as 90\kmsMpc\ (Grogin \&
Narayan 1996).

PG1115+080 has been observed by Schechter et al. (1997) who
derived a time delay and estimated a Hubble constant of 45\kmsMpc\
based on image and lens positions from Kristian et al. (1993) and
a lens redshift of about $z \approx 0.3$ from Henry \& Heasley (1986)
and Angonin-Willaime et al. (1993).  The observations reported here
were recently anticipated by Kundi\'c, Cohen, and Blandford
(1997, KCB) who find
$z = 0.311$, and use this to derive $H_0 = 62\pm17$.

B1422+231 is a very well studied QSO and Ly-$\alpha$ system at a
redshift of 3.62 discovered by Patnaik et al. (1992), with a lensing galaxy
at a redshift of $z=0.647$ according to Hammer et al. (1995).  There
is again a small group of three galaxies within a few arcseconds with
unknown redshift.  Although no time delay has been measured in this
system, the photometry compiled by Keeton and Kochanek (1996) suggest
that there may be enough variability in the QSO that it should be possible
to do so.

This article reports the first results in an ongoing program to
measure the most basic information necessary to exploit the
gravitational lenses: the redshift of the lensing galaxy, redshifts of
galaxies clustered around the lens, and velocity dispersions of the
stars within the galaxies.  Although a small ingredient compared
to measuring time delays and modelling potentials, it is an essential
one.

\section{Observations and Reductions}

B1422+231 and PG1115+080 were observed on March 30 and March 31, 1997
using the Low Resolution Imaging Spectrograph (LRIS) (Oke et al. 1995)
at the Keck II telescope on Mauna Kea, along with calibration
exposures, an observation of the cluster MS1358+62 to act as a
velocity dispersion calibrator, and HD132737 and AGK2+14873 as radial
velocity templates.  The observations are summarized in Table 1.  The
sky was clear and the seeing was about 0.8\arcsec\ throughout both
nights.  Long slits of 1.0\arcsec\ and 0.7\arcsec\ were used along
with gratings of 300 and 600~l/mm, both blazed at 5000\AA.  The
300~l/mm grating was always rotated to provide coverage from
3800--8700\AA, whereas the 600~l/mm grating was rotated either to
cover 4800--7680\AA\ for the QSO observations or 3800--6080\AA\ for
the templates.  The spectral resolution was about 7.9\AA\ \FWHM\ for
the 300~l/mm grating, 4.65\AA\ \FWHM\ for the 600~l/mm grating with a
1.0\arcsec\ slit, 3.59\AA\ \FWHM\ for the 600~l/mm grating with a
0.7\arcsec\ slit,  and the scale along the slit was
0.211\arcsec/pixel.  The template stars were guided smoothly across
the slit in many locations in order to have uniform illumination
across the slit, to build up signal to noise, and to map out loci of
constant slit position across the detector.  The slit was rotated as
illustrated in Figure 1, either to observe two companion galaxies
simultaneously or else to cover the lensing galaxy while avoiding as
much QSO light as possible.

\insfig{
\begin{figure}
\epsscale{1.0}
\plotone{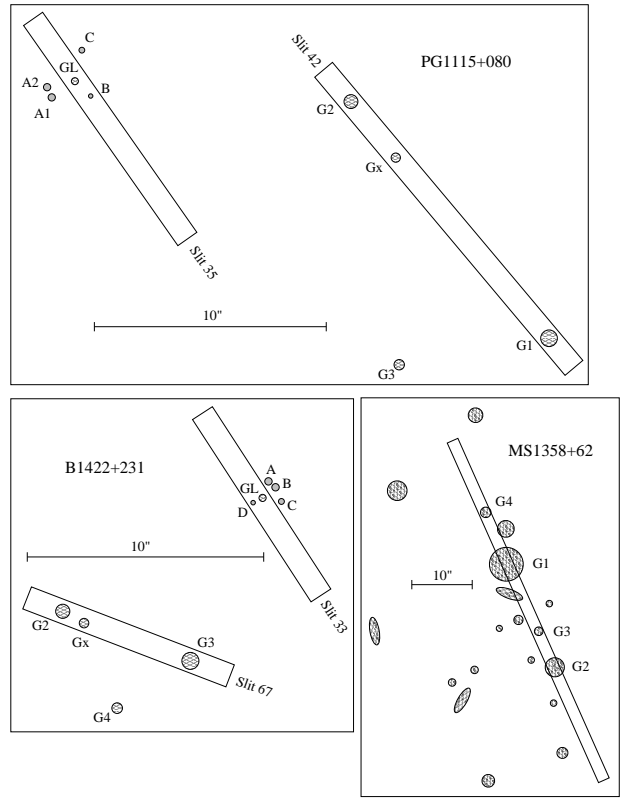}
\caption[fig1.eps]{
Illustration of the slit positions and galaxy idenifications for
B1422+231, PG1115+080, and MS1358+62.  North is up and East is left
for each diagram.
\label{fig1}}
\end{figure}
}

The spectra were reduced using software described in detail by Tonry
(1984).  The basic steps are to flatten the images, remove cosmic
rays, derive a wavelength solution as a function of both row and
column using sky lines (wavelengths tabulated by Osterbrock et
al. 1996), derive a slit position solution as a function of both row
and column using the positions of the template star images in the
slit, rebin the entire image to coordinates of log wavelength and slit
position, add images, and then sky subtract.  A quadratic fit to
patches of sky on either side of the object (including a patch between
for the galaxy pair observations) did a very good job of removing the
sky lines from the spectra.

As has been stressed by Kelson et al. (1997), measuring the velocity
dispersion of a galaxy at a redshift significantly larger than zero
must be done carefully, since the instrumental resolution of a
spectrograph tends to a constant number of angstroms, whereas the
redshifted spectrum has been stretched.  Hence a straightforward
cross-correlation or Fourier quotient will underestimate the
dispersion of the galaxy.  Kelson et al. measured dispersions in the
cluster MS1358+62 using high resolution template spectra and very
careful modelling of the spectrograph resolution derived by measuring
sky line widths.  The observations here use the trick that the ratio
between an 0.7\arcsec\ and 1.0\arcsec\ slit is slightly larger than
$(1+z)$ for these lenses, hence a {\it template} measured with the
0.7\arcsec\ slit will have almost exactly the same instrumental
resolution as a {\it galaxy} at a redshift of $z\approx0.3$.  This is
borne out by the ratio of the measured spectral resolutions: 
4.65\AA\ $\div$ 3.59\AA\ = 1.30.  This is applicable only to the
600~l/mm observations, of course.

As a test that this observing procedure will give correct dispersions,
we also observed the cluster MS1358+62 for which Kelson et al. measure a
dispersion of $305\pm5$ for the central galaxy, G1, and $213\pm4$
for the neighboring bright galaxy, G2 (Franx, private communication).

The analysis of the companion galaxies was straightforward since the
spectra were of very high signal to noise and were uncontaminated by
QSO light.  In each case the spectrum was extracted, and
cross-correlated with the template spectrum according to Tonry and
Davis (1979) as well as being analyzed by the Fourier quotient method
of Sargent et al. (1977).  (The cross-correlation is more robust in
the case of low signal to noise, but at the signal levels here the two
results are statistically the same, and are simply averaged.)  For
each spectrum the redshift, error, and velocity dispersion were
calculated.  The redshift calculation included the entire spectrum,
whereas the spectrum blueward of 4000\AA\ (rest frame) was excised for
the dispersion calculation, since the calcium H and K lines are so
broad that they are always problematic for dispersions.  Dispersions
are not shown for 300~l/mm observations, and they are not corrected
for any aperture effects, hence correspond to an aperture of
approximately 1\arcsec\ square.

Table 2 lists the redshifts, errors, velocity dispersions, errors, and
cross-correlation significance ``$r$'' values for each spectrum.

\subsection{B1422}

We were surprised to find an unmistakable emission line system
adjacent to B1422-G2, offset by about 1.0\arcsec\ along the line
towards B1422-G3.  It showed very clear, extended emission lines of
3727 and 5007 in both exposures, and H$\beta$ could also be faintly
discerned.  Lacking any imaging information on this galaxy, we
tabulate it as ``B1422-Gx''.

The lensing galaxy extractions require considerable care.  In the case of
B1422 we exploited the fact that the A-B-C components of the QSO would
put a more extended distribution of light along the slit than the
lensing galaxy.  We therefore extracted three swaths of approximately
1\arcsec; the central one we dubbed ``QSO+galaxy'' and the two
flanking swaths we added and dubbed ``QSO''.  Of course there is
considerable galaxy light in the flanking swaths but the expectation
was that the proportion of galaxy light would be less.  
We found a linear function of wavelength which, when multiplied
by the ``QSO'' spectrum, would match the Ly-$\alpha$ and C~IV QSO
lines in the ``QSO+galaxy'' spectrum.  Subtraction left behind a residue
which was a clean enough lensing galaxy spectrum that we could derive
a solid cross-correlation redshift.  

In principle it would have been better to get a pure QSO spectrum for
subtraction, and indeed we obtained a pure QSO spectrum by shifting
the slit down on top of the A--C line.  However, the QSO light in the
composite spectrum has a large gradient across the slit, hence a
wavelength shift and large intensity variations.  Time did not permit
rotating the spectrograph 180 degrees and offsetting to the other side
of the QSO to try to duplicate the QSO contribution to the composite
spectrum.

We see no emission in the B1422 lensing galaxy whatsoever, despite the
claims by Hammer et al. (1995) that 3727 and 5007 appeared at 6138\AA\
and 8247\AA.  We excised the A and B bands at 7625\AA\ and 6885\AA\
from the spectrum and 130\AA\ centered on the Ly-$\alpha$ peak, and
performed a cross-correlation against the template star.  The
cross-correlation came out with a correlation peak at $z = 0.3366$
with an $r$ value of 4.2, which is significant enough that it is quite
unlikely (probability around 1 percent) to be spurious.  It is
possible to see H+K, Mg, and Na at that redshift.  With the
corroboration from the neighboring galaxies, we are quite confident
that this is indeed the redshift of the lensing galaxy.

The spectra of the QSO, lensing galaxy, G3, G2, and Gx are plotted in
Figure 2.  The most prominent Fraunhofer lines are labeled for the
lensing galaxy, and while they can plausibly be seen by eye,
our trust in this redshift ultimately depends on the
cross-correlation significance.

\insfig{
\begin{figure}
\epsscale{1.0}
\plotone{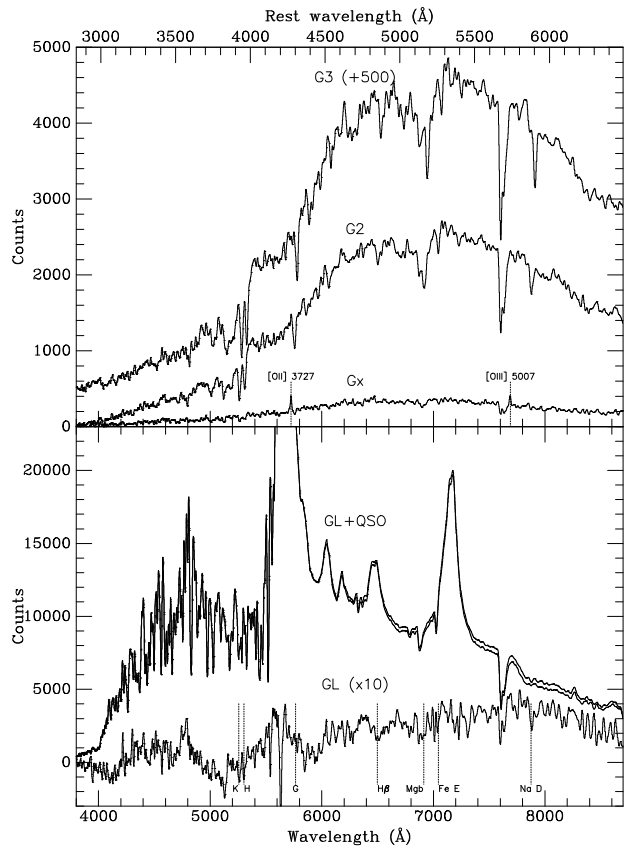}
\caption[fig2.eps]{ The spectra are shown for G3 (offset by 500 for
clarity), G2, Gx, QSO+lensing galaxy, ``pure'' QSO, and extracted
lensing galaxy (scaled by a factor of 10 for visibility) in the B1422
system.  These spectra have been Gaussian smoothed with a \FWHM\ of
6\AA.  The top axis shows rest wavelength at a redshift of 0.34.  The
[OII] and [OIII] lines in Gx are labeled as well as the Fraunhofer
lines in GL.
\label{fig2}}
\end{figure}
}

\subsection{PG1115}

As with B1422, we serendipitously picked up an emission line galaxy in
the PG1115 system along the slit between G1 and G2, lying 3.8\arcsec\
from G2.  It displays emission lines of 5007, 4959, H$_\beta$, and
H$_\gamma$, and we refer to it in Table 2 as ``PG1115-Gx''.

The slit for PG1115 ran over the top of the B QSO component, the
lensing galaxy, and skirted the edge of the C component, creating a
bimodal distribution of light in the slit with the galaxy in the
center.  In this case the QSO light illuminated the slit uniformly
enough that the subsequent exposure centered on the A1-A2 components
matched the QSO light from the B and C to a high degree of accuracy.
The B and C peaks were separated by 1.8\arcsec\ at this position
angle, and we extracted the inner 1\arcsec\ from between the two peaks
and subtracted a scaled version of the A1-A2 spectrum to give us the
lensing galaxy spectrum illustrated in Figure 3.

Comparison with the redshifts presented by Kundi\'c et al. shows
extremely good agreement: GL -- 0.3100 KCB, 0.3098 here; G1 -- 0.3099
KCB; 0.3098 here; G2 -- 0.3120 KCB; 0.3123 here. 
As noted by KCB, these redshifts are in good agreement
with Henry \& Heasley (1986), but are only marginally consistent with
Angonin-Willaime at al. (1993).

The spectra of the QSO, lensing galaxy, G1, G2, and Gx are plotted in
Figure 3.

\insfig{
\begin{figure}
\epsscale{1.0}
\plotone{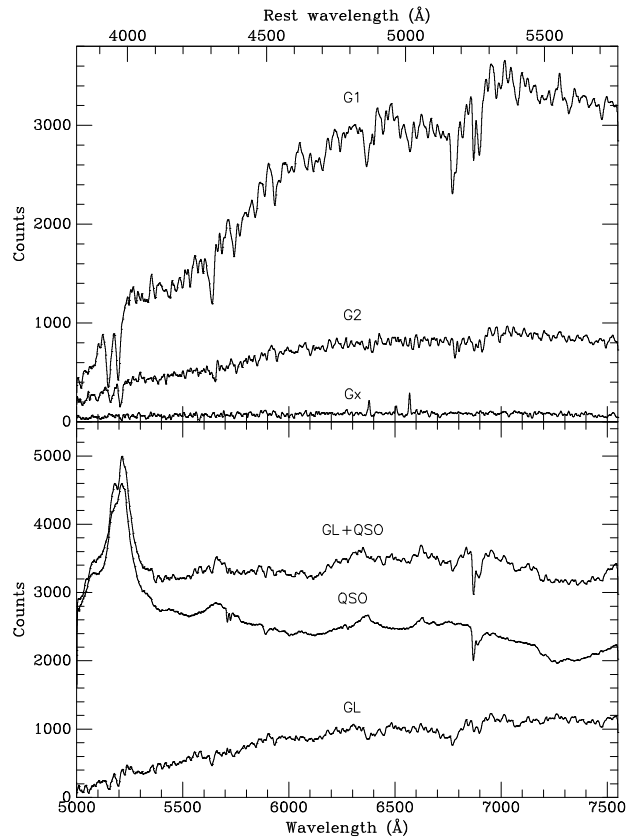}
\caption[fig3.eps]{ The spectra are shown for G1, G2, Gx, QSO+lensing
galaxy, pure QSO, and extracted lensing galaxy in the PG1115 system.
These spectra have been Gaussian smoothed with a \FWHM\ of 6\AA.  The
top axis shows rest wavelength at a redshift of 0.31.  Emission lines
of 5007, 4959, and 4861 are apparent in the spectrum of Gx.
\label{fig3}}
\end{figure}
}

\subsection{MS1358}

A pair of exposures of MS1358 identical to those of PG1115 yielded
spectra of seven galaxies.  The slit was aligned between the central
galaxy, called G1 here, and the bright galaxy 19\arcsec\ to the SW (PA
204), called G2.  The galaxies G3 and G4 were well centered in the
slit, hence their velocities and dispersions should be reliable.
The three remaining galaxies were marginally caught by the slit and
low signal-to-noise, so we do not report their velocities.
Comparison with Fabricant et al. (1991) for the three galaxies in
common (G1 = \#140, G2 = \#121, and G3 = \#126) gives a mean offset of
$\delta z = 0.0002$ and a scatter of $\sigma_z = 0.0001$, well within
the error estimates.

The comparison of velocity dispersions with Kelson et al. reveals no
significant offset in the dispersions measured here --- G1: $305\pm5$
versus $299\pm22$ here; G2: $213\pm4$ versus $235\pm23$ here.  The
mean of these ratios differs from unity by only 4\%,
indicating that the dispersions and errors reported here should be
accurate.

\section{Discussion}

Computing means and standard deviations for the redshifts presented
here, we find the B1422 lens is part of a group at 
$z_g = 0.339\pm0.002$ with a rest frame dispersion (i.e. standard
deviation of $cz$ divided by $1+z$) of 733\kms\ (3 galaxies).  The
PG1115 lens is part of a group at $z_g = 0.311\pm0.001$ with a rest
frame dispersion of 326\kms\ (5 galaxies).  Kundi\'c et al.  gave
165\kms\ but this is obviously a typographical error which should be
270\kms, (confirmed by Kundi\'c, private communication).  Using the
four galaxies observed in the MS1358 cluster we find 
$z_c = 0.325\pm0.001$ with a rest frame dispersion of 432\kms\ (4 galaxies).

A few comments seem appropriate.  The B1422 group appears to be quite
massive, particularly given the number of galaxies present.  Deeper
imaging would probably be worthwhile.  Despite the comparison Kundi\'c
et al. make between the PG1115 and Hickson poor groups (Hickson et
al. 1992), it now looks as though the group is much more massive, and
the dispersion of 326\kms\ is in quite good agreement with the models
of Schechter et al., who need a group dispersion of 383\kms\ to
provide the necessary shear to explain the image positions.  Finally,
the dispersion for MS1358 serves as a warning that small numbers of
galaxies near the centers of groups may not be a reliable measure of
the overall group mass.  Fabricant et al. determine a dispersion of
1125\kms\ for MS1358 from approximately 65 galaxies within 2
arcminutes of the BCG.  Recomputing a dispersion from their redshifts
but restricting to galaxies closer to the BCG than G2 we find a
dispersion of 625\kms, demonstrating how a cool subsystem can exist in
a massive cluster.

The redshift found here for the B1422 system is significantly lower
than that reported by Hammer et al., so the lens and companions are
not as highly luminous as had been suggested.  This lower
redshift is in quite good agreement with the redshift estimated
photometrically by Impey et al. (1996).  The rather high cluster mass
called for by Hogg and Blandford (1994) appears to be confirmed here.
The presence of the background galaxy at $z = 0.536$ at a distance of
9\arcsec\ from the lens may point to a background cluster which is
providing significant multiple lensing.  Very deep imaging of this
source would help elucidate whether there is indeed a background
cluster present.

The dispersion for the lensing galaxy in PG1115 of 281\kms\ can be
corrected according to the formula given by Jorgensen et al. 1995 to
the standard metric aperture commonly used in fundamental plane
studies: 3.4\arcsec\ at the distance of Coma.  This corrected
dispersion is 293\kms\ and it seems surprisingly large.  Perhaps the
best way to see this is to consider the Einstein ring created by an
isothermal lens of dispersion $\sigma$ aligned with the source.  This
ring will have radius $r_0$ given by
\begin{equation}
r_0 = {D_{ds} D_d \over D_s} {4\pi\sigma^2 \over c^2}.
\end{equation}
Using redshifts of $z_s = 1.722$, $z_d = 0.311$, and $\Omega_0 = 1$,
we find the first term is $394 h^{-1}$~Mpc, and the angular size
distance of the lens is $D_d = 580 h^{-1}$~Mpc, so that one arcsecond
is $2.81 h^{-1}$~kpc.  This then gives $r_0 = 4.72 h^{-1}$~kpc =
1.68\arcsec.  However, the geometric mean of the four QSO positions
around the lensing galaxy is 1.17\arcsec, which is the reason that the
models of Schechter et al. prefer a dispersion for the lens of about
235\kms.  The position of the ring of images is quite robust to the
details of shear and azimuthal image position, so it will take
detailed modelling to understand how this high dispersion can fit in
with the rest of the observables.  It
is conceivable that this dispersion is simply wrong, but the spectrum
is quite high quality, even given the necessity for QSO subtraction,
having a signal to noise of about 20 per angstrom.  It is also
possible that the dispersion drops rapidly from the measured value of
281\kms\ to something nearer 235\kms\ at the radius of 
$3.3 h^{-1}$~kpc where the images lie.  While not unheard of, this
would be a surprisingly fast decline in velocity dispersion.  While
the prospects for spatial resolution in the dispersion are not good,
it can be used to constrain possible models for the galaxy: an
isothermal mass distribution looks problematic.  The models of Keeton
and Kochanek (1997) explore some of these possibilities.  A final
possibility is that the shear from cluster is pushing the images
closer together, overcoming some of the splitting from the lens
itself, but it is unclear whether such a model could be workable.

In recent years the sophistication of gravitational lens models has
clearly outstripped the quality of the data available.  The Keck
telescopes and the LRIS spectrograph are truly marvelous facilities,
and it is the aim of this and subsequent papers to try to bring some
grist to the hungry mill, and perhaps help teach us about the
cosmology of our universe.

\acknowledgements
Thanks are due to Paul Schechter for encouragement and helpful
suggestions throughout this project.

\clearpage

\begin{deluxetable}{rlrrrrrrr}
\tablecaption{Observing Log.\label{tbl1}}
\tablewidth{0pt}
\tablehead{
\colhead{Obs\#} &  \colhead{Objects} & \colhead{UT Date}  &
\colhead{UT}  & \colhead{$\sec\,z$} & 
\colhead{PA} & \colhead{Exposure} & 
\colhead{Slit} & \colhead{Grating}
} 
\startdata
 1. & AGK2+14783    & 3/30 &  5:46 & 1.01 & 90 &       & 0.7 & 600/5000 \nl
 2. & PG1115 GL+B   & 3/30 &  8:55 & 1.02 & 35 &  1500 & 1.0 & 600/5000 \nl
 3. & PG1115 GL+B   & 3/30 &  9:22 & 1.02 & 35 &  1500 & 1.0 & 600/5000 \nl
 4. & PG1115 A1+A1  & 3/30 &  9:53 & 1.04 & 35 &   500 & 1.0 & 600/5000 \nl
 5. & PG1115 G1+G2  & 3/30 & 10:10 & 1.05 & 42 &  1500 & 1.0 & 600/5000 \nl
 6. & PG1115 G1+G2  & 3/30 & 10:36 &      & 42 &  1500 & 1.0 & 600/5000 \nl
 7. & MS1358 G1+... & 3/30 & 11:12 & 1.37 & 24 &  1500 & 1.0 & 600/5000 \nl
 8. & MS1358 G1+... & 3/30 & 11:39 & 1.36 & 24 &  1500 & 1.0 & 600/5000 \nl
 9. & B1422  G2+G3  & 3/30 & 12:21 & 1.00 & 67 &  1500 & 1.0 & 300/5000 \nl
10. & B1422  G2+G3  & 3/30 & 12:47 &      & 67 &  1500 & 1.0 & 300/5000 \nl
11. & B1422  GL     & 3/30 & 13:54 & 1.09 & 33 &  1500 & 1.0 & 300/5000 \nl
12. & B1422  GL     & 3/30 & 14:21 & 1.15 & 33 &  1500 & 1.0 & 300/5000 \nl
13. & HD132737      & 3/30 & 15:11 & 1.20 & 90 &       & 1.0 & 300/5000 \nl
14. & HD132737      & 3/30 & 15:23 &      & 90 &       & 0.7 & 600/5000 \nl
15. & AGK2+14783    & 3/31 &  5:26 & 1.01 & 90 &       & 0.7 & 600/5000 \nl
16. & AGK2+14783    & 3/31 &  5:45 & 1.01 & 90 &       & 1.0 & 300/5000 \nl
17. & PG1115 GL+B   & 3/31 &  8:11 & 1.04 & 35 &  1500 & 0.7 & 300/5000 \nl
18. & PG1115 GL+B   & 3/31 &  8:35 & 1.02 & 35 &  1500 & 0.7 & 300/5000 \nl
19. & PG1115 A1+A2  & 3/31 &  9:05 & 1.02 & 35 &   500 & 1.0 & 300/5000 \nl
\enddata
\tablecomments{PA is east from north, exposures are in seconds,
slit widths are in arcseconds.}
\end{deluxetable}

\begin{deluxetable}{lrrrrrr}
\tablecaption{Redshifts and Dispersions.\label{tbl2}}
\tablewidth{0pt}
\tablehead{
\colhead{Galaxy} & \colhead{$y$} & \colhead{$z$} & \colhead{$\pm$}  &
\colhead{$\sigma$}  & \colhead{$\pm$} &  \colhead{$r$}
} 
\startdata
B1422-GL  &       & 0.3366 & 0.0004 &      &    &  4.2 \nl
B1422-G2  & $-2.9$& 0.3376 & 0.0001 &      &    & 13.0 \nl
B1422-G3  & $+2.9$& 0.3427 & 0.0001 &      &    & 11.4 \nl
B1422-Gx  & $-1.9$& 0.5360 & 0.0005 &      &    &  em. \nl
PG1115-GL &       & 0.3098 & 0.0002 &  281 & 25 & 10.4 \nl
PG1115-G1 & $-6.7$& 0.3098 & 0.0001 &  256 & 20 & 14.8 \nl
PG1115-G2 & $+6.7$& 0.3123 & 0.0001 &  130 & 60 & 10.9 \nl
PG1115-Gx & $+2.9$& 0.3121 & 0.0002 &      &    &  em. \nl
PG1115-GL &       & 0.3095 & 0.0002 &      &    &  8.1 \nl
MS1358-G1 & $+9.6$& 0.3272 & 0.0001 &  299 & 22 & 11.9 \nl
MS1358-G2 & $-9.5$& 0.3235 & 0.0001 &  235 & 23 & 14.1 \nl
MS1358-G3 & $-1.7$& 0.3248 & 0.0001 &  155 & 56 &  9.4 \nl
MS1358-G4 &$+18.3$& 0.3229 & 0.0002 &  175 & 86 &  5.2 \nl
\enddata
\tablecomments{Columns: 
Galaxy name, slit position (\arcsec), redshift and error, velocity
dispersion (\kms) and error, cross-correlation $r$ value.}
\end{deluxetable}

\endinsfig

\clearpage

\centerline{\bf FIGURE CAPTIONS}
\bigskip

\figcaption[fig1.eps]{
Illustration of the slit positions and galaxy idenifications for
B1422+231, PG1115+080, and MS1358+62.  North is up and East is left
for each diagram.
\label{fig1}}

\figcaption[fig2.eps]{ The spectra are shown for G3 (offset by 500 for
clarity), G2, Gx, QSO+lensing galaxy, ``pure'' QSO, and extracted
lensing galaxy (scaled by a factor of 10 for visibility) in the B1422
system.  These spectra have been Gaussian smoothed with a \FWHM\ of
6\AA.  The top axis shows rest wavelength at a redshift of 0.34.  The
[OII] and [OIII] lines in Gx are labeled as well as the Fraunhofer
lines in GL.
\label{fig2}}

\figcaption[fig3.eps]{ The spectra are shown for G1, G2, Gx, QSO+lensing
galaxy, pure QSO, and extracted lensing galaxy in the PG1115 system.
These spectra have been Gaussian smoothed with a \FWHM\ of 6\AA.  The
top axis shows rest wavelength at a redshift of 0.31.  Emission lines
of 5007, 4959, and 4861 are apparent in the spectrum of Gx.
\label{fig3}}

\end{document}